\documentclass[aps,pra,twocolumn,showpacs]{revtex4}
\usepackage{subfigure}
\usepackage{graphicx}
\usepackage{dcolumn}
\usepackage{amsmath}
\usepackage{amssymb}
\usepackage{multirow}
\usepackage{float}
\usepackage{epstopdf}
\usepackage[dvips]{color}
\usepackage{bm}

\def\beq{\begin{equation}}
\def\eeq{\end{equation}}

\begin{document}
\title{The two-atom energy spectrum in a harmonic trap near a Feshbach
resonance at higher partial waves }
\author{Akira Suzuki}
\affiliation{Department of Physics, Tokyo University of Science,
 Tokyo, Japan 162-8601}
\author{Yi Liang}
\affiliation{Department of Physics and Astronomy,
McMaster University, Hamilton, ON, Canada L8S 4M1}
\author{Rajat K. Bhaduri}
\affiliation{Department of Physics and Astronomy,
McMaster University, Hamilton, ON, Canada L8S 4M1}

\date{\today}

\begin{abstract}
Two atoms in an optical lattice may be made to interact strongly
at higher partial waves near a Feshbach resonance. These atoms, under
appropriate constraints, could be bosonic or fermionic. The universal
$l=2$ energy spectrum for such a system, with a caveat, 
is presented in this paper, and checked with the spectrum obtained by  
direct numerical integration of the Schr\"odinger equation. 
The results reported here extend
those of Yip for $p$-wave resonance (Phys. Rev. A {\bf 78}, 013612
(2008)), while exploring the limitations of a universal expression for 
the spectrum for the higher partial waves.
\end{abstract}
\maketitle

\section{Introduction}
Normally, the scattering between two neutral atoms in higher partial waves
is suppressed due to the centrifugal barrier. The situation may
change, however, by sweeping a static magnetic field near a Feshbach
resonance (FR) at a higher partial wave~\cite{regal, marte}.
In such a situation, the relative energy of the scattering atoms is at
near-coincidence with the energy of a quasi-bound molecular state with
a nonzero angular momentum. FR was first observed with bosons
interacting in the $s$-state \cite{ino}, and later in higher partial
waves~\cite{marte}. Two spin polarized fermions may interact in
a relative odd-$l$ state in a single channel ultra-cold gas, or in an even-$l$
state when the two atoms are in distinct spin states. The
single-channel $p$-wave FR was first observed by Regal {et
  al.}~\cite{regal} between $^{40}K$ atoms, while the fermionic $s$-wave
between these atoms in distinct spin states was observed by the same
group a little earlier~\cite{loftus}.

The two-particle energy spectrum in a spherical harmonic oscillator
(HO) near a $s$-wave FR is found to be universal when the range of the
interaction between the two atoms is much smaller than the
average interparticle distance in the HO, i.e. the oscillator length ~
\cite{busch, jonsell, brandon}. Near the Feshbach
resonance, this spectrum has been checked experimentally~\cite{stof}
in the limit of low tunneling for fermionic $^{40}K$ atoms in an
optical trap in two distinct spin
states. This is remarkable because the theoretical spectrum
is obtained in a one-channel approximation, and fits the data even with
the first term in the effective range expansion.
More recently, Yip~\cite{yip} has
obtained the spectrum of two identical fermions in a HO near a $p$-wave
FR. It is known that the $p$-wave FR gets split ~\cite{tick}
between $|m_l|=1$ and $0$. Yip's calculation is for $m_l=0$, but may be
generaised for nonzero $m_l$. For higher partial waves, however,
it was stated that the energy spectrum could not be expressed
in terms of the parameters in the
scattering amplitude and the oscillator constant~\cite{yip, yip1}. 
In this paper, using a method first formulated by
Jonsell~\cite{jonsell} for $l=0$, we express the energy spectrum for 
the higher partial waves in terms of the scattering parameters. We
examine the $l=2$ case in detail, and compare the spectra obtained 
from this analytical formula with the results obtained by 
a numerical integration of the Schr\"odinger equation. Yip's assertion
is justified, but only in a narrow region at resonance where the
ground state energy goes to zero. The analytical formula accurately
reproduces the excited state spectrum in the entire range, but is
shown to be not accurate for the ground state in a narrow band across
the resonance. With this exception, even though the energy spectra for $l=1$
and $l=2$ are extremely sensitive to the choice of the effective
range parameter, we find that the levels are practically unchanged
by introducing the next term in the expansion, which is
shape-dependent.

In calculating the $p$ wave spectrum, Yip used the effective
range expansion for higher partial waves~\cite{mott}
\begin{equation}
k^{2l+1}{\rm{cot}}\delta_l(k)=-\frac{1}{a_l}+\frac{1}{2} r_l k^2 ,
\label{range}
\end{equation}
with $l=1$. Note that the scattering length  $a_l$ and the effective range $r_l$
have the dimensions of $(L)^{2l+1}$ and $(L)^{-2l+1}$
respectively. For single channel elastic scattering by a power-law potential
$r^{-n}$, and {\it without any virtual transition to other channels like
an excited quasi-molecular state}, $a_l$ in Eq.(\ref{range}) is
only defined if $n>(2l+3)$~\cite{LL}. For the effective range $r_l$ to
exist, the restriction is even more severe, $n>(2l+5)$. Taking $n=6$ for the
asymptotic behaviour of the interatomic potential, we see that for $l=1$, only
$a_1$ exists, but not $r_1$. For $l=2$, neither $a_2$, nor $r_2$ is
defined~\cite{LL, gao}.

The situation changes, however, for dressed atoms near a Feshbach
resonance. Consider scattering in an open channel which is coupled
to a resonance in the closed channel through a spin-dependent
{\it{two-body}} interaction $W$ that depends on the relative
distance $r$ between the two atoms~\cite{kohler}. One part is the
long range tensor interaction between the two dipoles that falls off
as $r^{-3}$. The other is the spin-exchange interaction between the
two valence electrons of the alkali atoms that are closer than the
uncoupling distance $r_u$. For $r<r_u$, the nuclear and electronic
spins that were otherwise coupled in an isolated atom get uncoupled
due to one atom's proximity to the other atom. The resulting
spin-exchange interaction has a range $r_u$ which is smaller than
the so-called van der Waals distance, a measure of the length scale
for the $r^{-6}$ potential. The coupling potential $W$ may therefore
be regarded as of shorter range than $r^{-6}$ in cases where the
spin exchange interaction is dominant.

Elimination of the closed channel results in an additional effective
potential in the single-channel formalism. This effective potential,
which dominates the open channel scattering near the Feshbach
resonance, contains $W$ quadratically. The effective range expansion
should be valid for the higher partial waves for this short range
potential, as will be discuused in the next section.



\section{The two-body problem}
\subsection{Two-channel scattering}
We first recall the well-known results of two-channel scattering,
following the treatment and notation of
Cohen-Tannoudji~\cite{tannoudji}. The Hamiltonian $H$ is a $2\times
2$ matrix, with the diagonal elements $H_{op}$ and $H_{cl}$, and the
nondiagonal coupling given by the short range spin exchange
potential $W$. Here $H_{op}=(-\nabla^2+V_{op})$, with $V_{op}$ given
by the van der Waal power law potential whose asymptotic fall-off
goes like $r^{-6}$. The closed channel Hamiltonian, in the
single-resonance approximation, may be expressed as $H_{cl}=E_{res}
|\phi_{res}><\phi_{res}|$, with the resonance energy at $E_{res}$.

The wave function $\psi$ is a column matrix with two components
$\phi_{op}$ and $\phi_{cl}$. It is straightforward to show that the
two-channel scattering problem may be described entirely in the open
channel by a Lippman-Schwinger equation in which the incident wave
is distorted by the open channel potential $V_{op}$, and the
scattering is by the effective potential $V_{eff}$. More explicitly,
the (relative) outgoing wave of the dressed atoms $|\phi_{op}^k>$,
in terms of the distorted wave$|\phi_{k}^{+}> $, is given by
\beq
|\phi_{op}^k>=|\phi_k^{+}>+G_{op}^{+}(E) V_{eff}|\phi_k^{+}>~,
\label{wfn}
\eeq
where
\beq
V_{eff}=W\frac{|\phi_{res}><\phi_{res}|}{E-E_{res}-<\phi_{res}
|WG_{op}^{+}(E)W|\phi_{res}>}W~. \label{effective} \eeq In the
above, $G_{op}^{+}(E)=(E-H_{op}+i\epsilon)^{-1}$. The distorted
incident wave $|\phi_k^{+}>$ itself obeys a Lippman-Schwinger
equation in which the incident wave is a plane wave, the Green's
function is $(E-T+i\epsilon)^{-1}$, and the scattering potential is
$V_{op}$. As $k\rightarrow 0$, the large-$r$ behaviour of the first
term on the RHS of Eq.(\ref{wfn}) gives the background scattering
length, and the second term yields the energy-dependent scattering
length that dominates near the Feshbach resonance. Since we have in
mind partial waves $l\geq 2$, and $V_{op}(r)=-(\hbar^2/M)C_6/r^6$, the
background scattering length does not exist~\cite{LL, gao}. Indeed,
the phase shift caused by this potential is vanishingly small for
small $k$. This background phase shift, denoted by
$\delta^{bg}_l(k)$, is given by~\cite{levy}
\beq \tan\delta^{bg}_l(k)=\frac {\pi}{2} 2^{-5} C_6 \frac{\Gamma(5)
\Gamma(l-3/2)}{\Gamma^2(3)\Gamma(l+7/2)} ~|k|^3 k~.
\eeq
The distortion in the incident wave $\phi_k^{+}$ for small $k$ may
therefore be neglected, and the scattering for the higher partial
waves at low energies is governed by the energy-dependent short
range potential $V_{eff}$. As usual, the energy denominator in it
may be related to the Zeeman splitting due to a sweeping magnetic
field, giving rise to a large variation of the scattering length
$a_l$ near FR. We may then express the flow of the energy levels as
a function of the scattering length, assuming a fixed effective
range $r_l$ in the expansion (\ref{range}). This we proceed to do,
generalising a method first proposed by Jonsell~\cite{jonsell} for
$l=0$. We confine our treatment to $m_l=0$.

\subsection{Energy spectrum in harmonic trapping}
Each particle, of mass $M$,
moves in a harmonic potential $(1/2)M\omega^2 r^2$. Making the usual
transfomations to relative and CM co-ordinates, ${\bf r}=({\bf
 r}_1-{\bf r}_2)$, and $ {\bf R}=({\bf r}_1+{\bf r}_2)/2$, and their
corresponding canonical momenta ${\bf p}, {\bf P}$,
we obtain, for the noninteracting particles,
\beq
H_0=\left(\frac{P^2}{2M_{cm}}+\frac{1}{2} M_{cm}\omega^2R^2\right) + \left(\frac{p^2}
{2\mu}+\frac{1}{2} \mu \omega^2 r^2\right)~,
\eeq
where $M_{cm}=2M$, $\mu=M/2$. Consider the relative motion of these
two trapped particles, interacting with a short range effective
potential $V_s(r)$. This is the single-channel equivalent of the
potential $W^2(r)$ of Eq.(\ref{effective}), with the energy dependence
in the denominator being absorbed in the rapidly varying scattering
length parametrized by the Zeeman splitting. The two-body
Schr\"{o}dinger equation in each partial wave $l$
for the radial wave function $u_l(r)=r\psi_l(r)$ with energy $E_l$ may
be expressed in dimensionless variables
$x=r/(\sqrt{2}~L)$, where $L=\sqrt{\hbar/(M\omega)}$, and
$\eta_l=2E_l/({\hbar\omega})$. It is given by
\beq
-\frac{d^2 u_l}{dx^2}+\frac{l(l+1)}{x^2}u_l+\frac{2V_s}{\hbar\omega}
u_l+ x^2 u_l=\eta_l u_l~.
\label{wave}
\eeq
We may find the energy spectrum of the above equation without
specifying the specific form of $V_s$ by generalising a method first
adopted by Jonsell~\cite{jonsell} for $l=0$. Let the range of the
short range potential $V_s(r)$ be given by $b$. For $r>b$, taking
$V_s=0$, the solution of Eq.(\ref{wave}) is given by
\begin{widetext}
\begin{equation}
u_l={\rm e}^{-x^2/2}
\left[c'_1x^{l+1}M\left(\frac{2l+3-\eta_l}{4},l+\frac{3}{2};y\right)
+c'_2x^{-l}M\left(\frac{-2l+1-\eta_l}{4},\frac{1}{2}-l;y\right)\right] \;,
\label{35}
\end{equation}
\end{widetext}
where $y=x^2$, and $M(\alpha,\gamma;z)$ is the confluent
hypergeometric function~\cite {LL}.
Since, for large $z$, $M(\alpha,\gamma;z)$ behaves as
\begin{equation}
M(\alpha,\gamma;z)\sim \frac{\Gamma(\gamma)}{\Gamma(\alpha)}z^{\alpha-\gamma}
{\rm e}^{z} \;,
\end{equation}
the wave function behaves as
\begin{widetext}
\begin{equation}
\label{38}
u_l=\left[c'_1\frac{\Gamma(3/2+l)}{\Gamma((3+2l-\eta_l)/4)}
+c'_2\frac{\Gamma(1/2-l)}{\Gamma((1-2l-\eta_l)/4)}\right]
x^{-(1+\eta_l)/2}{\rm e}^{x^2/2}
\end{equation}
\end{widetext}
for large $x$.
In order to get a convergence solution, we have to have
\begin{equation}
\frac{c'_2}{c'_1}=-\frac{\Gamma(3/2+l)}{\Gamma(1/2-l)}
\frac{\Gamma((1-2l-\eta_l)/4)}{\Gamma((3+2l-\eta_l)/4)} \;.
\end{equation}
Since
\begin{equation}
\frac{\Gamma(3/2+l)}{\Gamma(1/2-l)}=(-1)^l\left(l+\frac{1}{2}\right)
\frac{[(2l-1)!!]^2}{2^{2l}} \;,
\end{equation}
we obtain
\begin{equation}
\label{40}
\frac{c'_2}{c'_1}=-(-1)^l\left(l+\frac{1}{2}\right)
\frac{[(2l-1)!!]^2}{2^{2l}}
\frac{\Gamma((1-2l-\eta_l)/4)}{\Gamma((3+2l-\eta_l)/4)} \;.
\end{equation}
We now need to relate this ratio to the scattering length $a_l$ to
determine the eigenvalue $\eta_l$. Note
that if the range $b$ of $V_s$ is very small, as we approach
$r \rightarrow b^{+}$, the oscillator potential may be neglected. {\it This
assumption is crucial for our derivation, and is examined in some
detail in Appendix A for $l=2$}. Thus
$u_l(r)$ for positive energy may be regarded as the phase-shifted scattering
solution due to $V_s$, given by
\begin{equation}
\label{31}
u_l(r)=A_l kr\left[j_l(kr)-n_l(kr)\tan\delta_l\right] \;,
\end{equation}
$E_l=\hbar^2k^2/M$. For $r>b$, but still $\simeq 0$, $u_l(r)$ behaves as
\begin{equation}
\label{32}
u_l(r)\;\mathop{\longrightarrow}\;
B_lr^{l+1}\left[1+(2l+1)[(2l-1)!!]^2\frac{\tan\delta_l}{(kr)^{2l+1}}\right] \;,
\end{equation}
where
\begin{equation}
B_l=\frac{A_lk^l}{(2l+1)!!}\;.
\end{equation}
We match $u_l$, given by Eq.(\ref{32}) with
that obtained from Eq.(\ref{35}) for small $x$.  The validity of this
procedure for $l=2$ rests on the condition (\ref{apex2}) derived in 
Appendix A. For $z\rightarrow 0$, and $\gamma\neq 0$,
the Kummer's functions $M(\alpha,\gamma;z)$ is
unity; hence the  solution given by Eq.(\ref{35}) behaves as
\begin{equation}
u_l(x)=c'_1x^{l+1}\left(1+\frac{c'_2}{c'_1}x^{-2l-1}\right) \;.
\label{harmo1}
\end{equation}
When this is matched with Eq.(\ref{32}), we obtain
\begin{equation}
\label{42}
\frac{c'_2}{c'_1}=(2l+1)[(2l-1)!!]^2\frac{\tan\delta_l}{(\sqrt{2}kL)^{2l+1}}\:.
\end{equation}
Equating Eqs(\ref{40},~\ref{42}), and using the effective range
expansion (\ref{range}), we obtain
\begin{equation}
\label{46}
\frac{1}{\sqrt{2}}\frac{\Gamma((1-2l-\eta_l)/4)}{\Gamma((3+2l-\eta_l)/4)}
=(-2)^l\frac{\tilde{a}_l}{1-\tilde{a}_l\tilde{r}_l\eta_l/4} \;.
\end{equation}
In the above, we have defined the dimensionless quantities
\begin{equation}
\label{45}
\tilde{a}_l=\frac{a_l}{L^{2l+1}}\;, \qquad \tilde{r}_l=L^{2l-1}r_l\;.
\end{equation}

If we set $l=1$ and $L=l_r/\sqrt{2}$ following Yip~\cite{yip},
we recover his result, that is, his Eq.(13);
\begin{equation}
-\frac{l_r^3}{v}+\frac{1}{2}l_r(2c)\eta_1
=8\frac{\Gamma((5-\eta_1)/4)}{\Gamma((-1-\eta_1)/4)}~ ,\
\label{main}
\end{equation}
where we have put $a_1=v$ and $r_1=2c$. Similarly, by setting $l=0$
in Eq.(\ref{46}) we recover the spectrum of the $l=0$ states as
given by Busch~\cite{busch}. Note that Eq.(\ref{46}) has been
obtained with no mention of any specific shape of the potential, and
is valid for {\it any} short-range two-body potential.\\

\section{ Results and Discussion }

In Figs 1-3, we plot the energy spectrum given by Eq.(\ref{46}) as a
function of the scaled scattering length $\tilde{a}_l$ for $l=0-2$,
keeping the effective range $\tilde{r}_l$ fixed. The same energy
spectra look quite different when plotted against the inverse of the
scattering length, $1/\tilde{a}_l$. For $l=1$, we reproduce Yip's
result, and do not duplicate it here. Our Fig. 4 shows the energy
levels for $l=2$, plotted as a function of $1/\tilde{a}_l$. Before
discussing these spectra, we comment on the choice of the effective
range parameter. For $l=1$, Yip had set the scaled effective range
$\tilde{r}_1=-64/\sqrt{2}$ from experimental data~\cite{tick}. From
a theoretical point of view, to
see if this may be generated by a potential whose range is much
smaller than the oscillator length $L$, we take a square-well
potential. The shape of the potential should not matter unless the
shape-dependent term in the effective range expansion (proportional
to $k^4$) affects the energy spectrum. We have verified that the
spectra in Figs 1-4 remain virtually unchanged when this term is
included. In the Appendix B, the analytical expressions for the
scattering length and the effective range for any given partial wave
are given. At a resonance, $a_l=\pm\infty$, hence it follows from
Eq.(\ref{A2}) that $j_{l-1}(s)=0$. Therefore, from Eq.(\ref{A3}) we
get $r_0=b$, $r_1=-3/b$, and $r_2=-15/b^3$. For Yip's choice of
$r_1L=\tilde{r_1}=-64/\sqrt(2)$, it follows that $b/L=\tilde{b}\simeq 1/15$,
fulfilling the condition that $b<<L$. Unlike the case $l=1$, we do not
have guidance from experiment for the choice of $\tilde{r_2}$, so
we deduce it from the square-well potential with $\tilde{b}=1/15$, $1/30$,
and $1/10$. Furthermore, we find that for the square-well example,
although the scattering length is highly
sensitive to the choice of the strength parameter $s$ defined by
Eq.(\ref{A1}), the effective range hardly changes as $s$ is varied
over a narrow range to accommodate the variation in the scattering
length shown in Figs.(1-4).
\begin{minipage}[t]{7.8cm}
\begin{figure}[H]
\begin{center}
\includegraphics[width=1.0\textwidth]{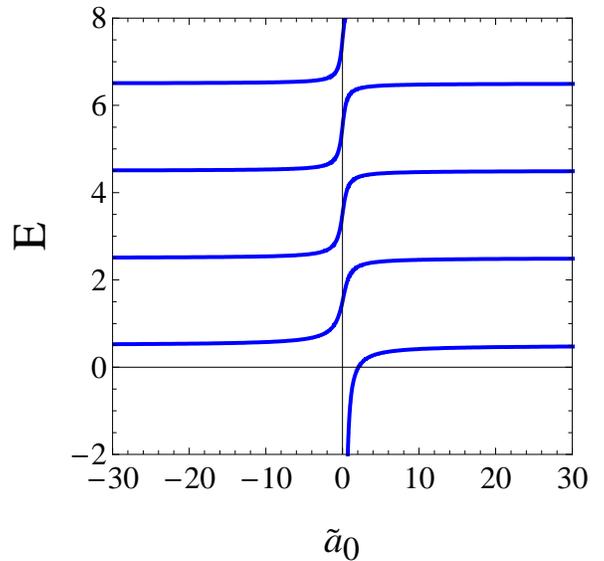}
\caption{(Color online) Plot of the $l=0$ energy levels $E$ in 
 units of $\hbar\omega$
versus the s-wave scattering length $a_0$ in units of the oscillator
length $L$. The scaled effective range is fixed at $1/15$. The plots
change negligibly even for zero range.}
\end{center}
\end{figure}
\end{minipage}
\hspace{0.5cm}
\begin{minipage}[t]{7.8cm}
\begin{figure}[H]
\begin{center}
\includegraphics[width=1.0\textwidth]{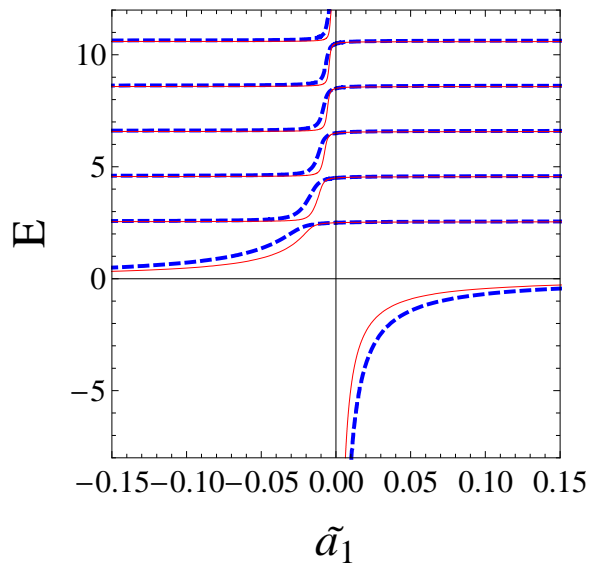}
\caption{(Color online) The~ $l=1$ energy levels vs the scaled 
$p$-wave scattering
length $\tilde{a}_1$. The dashed curves are for the effective
range $\tilde{r}_1=-30$, while the continuous curves are for
$\tilde{r_1}=-45$.}
\end{center}
\end{figure}
\end{minipage}
\begin{figure}[H]
\centering
\subfigure[]{\includegraphics[width=0.45\textwidth]{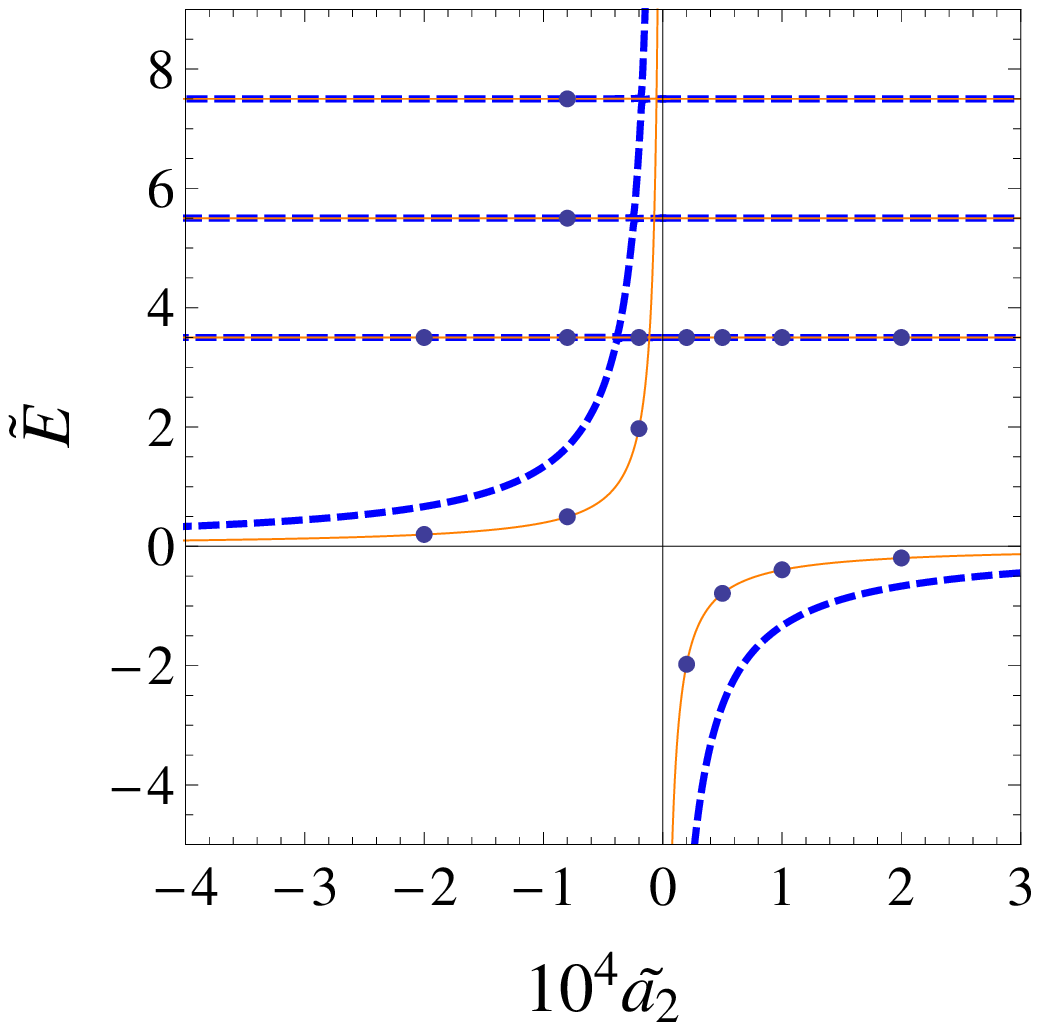}}
\subfigure[]{\includegraphics[width=0.45\textwidth]{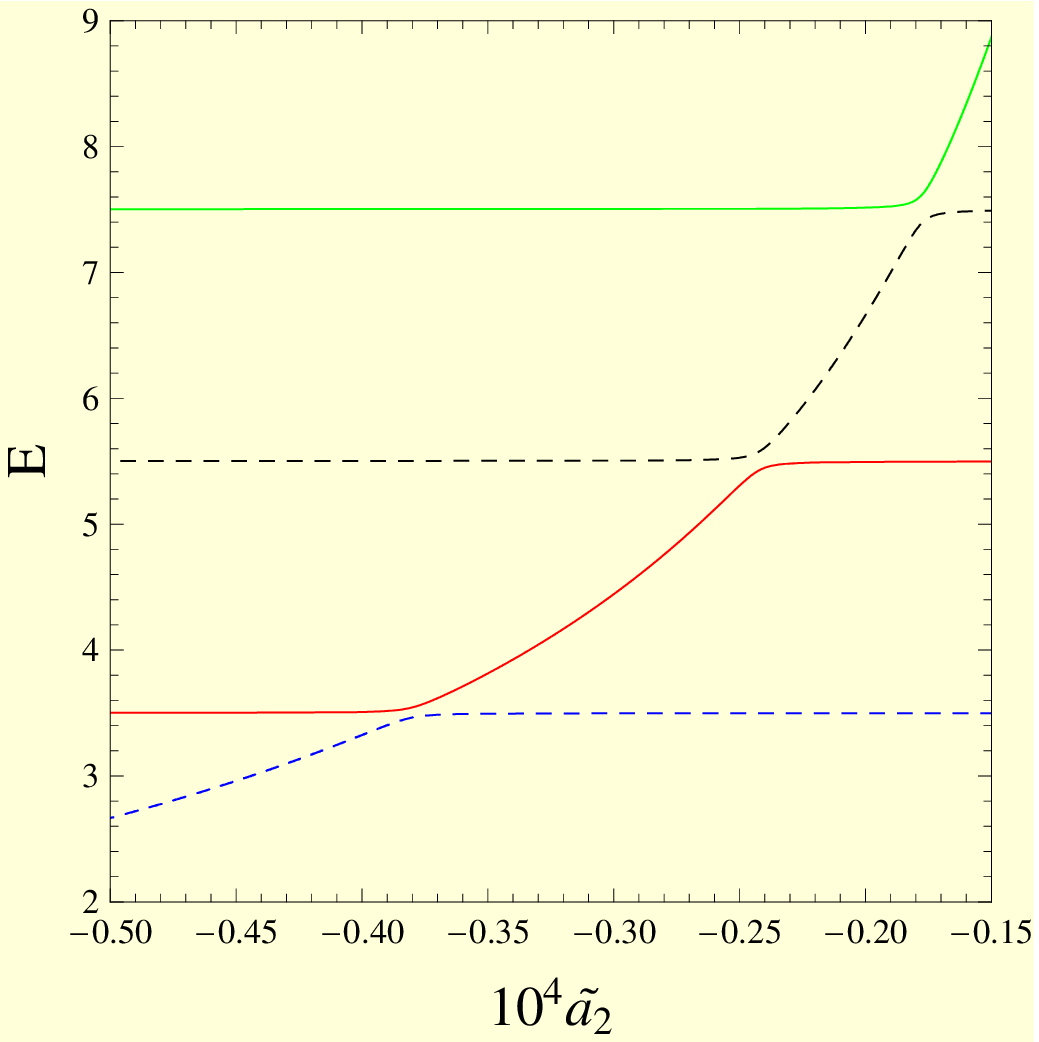}}
\renewcommand{\figurename}{FIG.}
\caption{(Color online) (a): The $l=2$ energy levels vs the scaled 
$d$-wave scattering length. The dashed line is for
$\tilde{r}_2=-15\times 10^3$, while the continuous line is for
$\tilde{r}_2=-15^{4}$. For the excited states, the dashed and
continuous lines cannot be distinguished on this scale.
See text for the choice of the effective range
parameters. The superimposed dots are the numerically calculated
energies obtained by solving the eigenvalue equation (\ref{wave})
directly. (b): Amplified version for the excited states showing that
the levels do not actually cross, but bend sharpely in Fig 3a.}
\label{}
\end{figure}

Figs. (1-3) show that away from the resonance
($\tilde{a}_l\rightarrow\pm\infty$), the energy plots against $\tilde{a}_l$ have
similar shapes. But whereas the $l=0$ plots remain essentially
unchanged when $\tilde{r_0}$ is varied over a wide range, the higher
partial waves become more and more sensitive to the choice of the
effective range. This is apparent from Figs.2-3. In all three, the
lowest energy state tends to $-\infty$ as $\tilde{a}_l\rightarrow 0$.
When $\tilde{a}_l\rightarrow \pm \infty$, the $l=0$ energy levels tend to
the limit $(2n+1/2), n=0,1,2..$ in units of the oscillator spacing. By
contrast, for $l\neq 0$, all but the lowest energy level go to the
noninteracting values $(2n+l+3/2)$ in the zero range limit. All
energies are in units of the oscillator spacing. At resonance, the
lowest energy level for $l\neq 0$ tends to zero as the range of the
potential is decreased.

\begin{figure}[H]
\begin{center}
\includegraphics[width=0.5\textwidth]{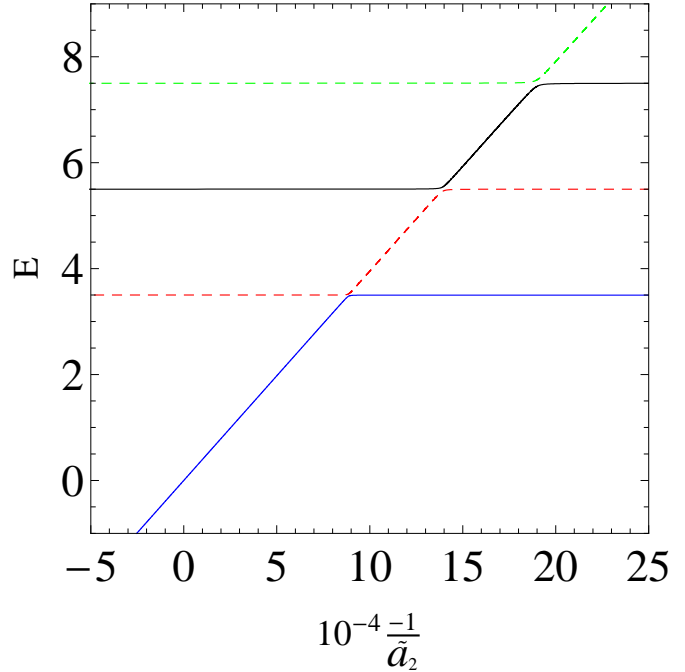}
\caption{(Color online) Plot of the excited state $l=2$ energy  
levels vs the inverse of the scattering length in dimensionless
units. This figure should be compared to Fig. 3 b.}
\end{center}
\end{figure}

In Fig. 4, we see that the higher energy levels show strong
bends away from the resonance, a feature even more pronounced than
for the $l=1$ case shown in~\cite{yip}.

\begin{table*}
\begin{center}
\caption{Comparison of energy $\eta_2$ obtained numerically 
from Eq.(\ref{wave}) and from the analytical formula (\ref{46}).
The range $\tilde{b})$ of the square-well potential is kept fixed at 
$1/15$}
\begin{ruledtabular}
\begin{tabular}{r|l|l|l|l|l|l|l|l|l|l|l|l|l}
$l=2$ & $10^4\widetilde{a_2}$ & & -2 & -0.8 & -0.2 & 0.2 & 0.5 & 1 & 2 & 100 & $10^5$ & $10^8$\\
\hline
\multirow{4}{*}{$\widetilde{E}$($\hbar \omega$)} & \multirow{2}{*}{Ground state} & N\footnote{Numeric results} & 0.1992 & 0.4952 & 1.9726 & -1.9772 & -0.7887 & -0.3945 & -0.1956 & -0.0022 & 0.0017 & 0.0017  \\  \cline{3-13}
& & A\footnote{Analytical results} & 0.1975 & 0.4939 & 1.9762 & -1.9731 & -0.7894 & -0.3951 & -0.1976 & -0.0040 & -0.0001 & -0.0001\\  \cline{2-13}
& \multirow{2}{*}{First excited state} & N & 3.5002 & 3.5010 & 3.5012 & 3.5001 & 3.5001 & 3.5001 & 3.5002 & 3.5002 & 3.5002  & 3.5002 \\ \cline{3-13}
& & A & 3.5002 & 3.5002 & 3.5003 & 3.5001 & 3.5001 & 3.5001 & 3.5002 & 3.5002 & 3.5002 & 3.5002 \\
\end{tabular}
\end{ruledtabular}
\end{center}
\end{table*}

To confirm that the analytical result (\ref{46}) does produce
the energy spectrum for $l=2$ accurately for the range of the
scattering parameters shown in Fig. 3(a), we solved Eq.(\ref{wave})
numerically to determine $\eta_2$. For $V_s$, a square-well potential
was taken with $\tilde{b}=1/15$, and it was kept fixed. The scattering
length $\tilde{a}_2$ was varied as shown by changing the 
depth of the potential. In Fig. 3(a), the dashed curves and lines are
plotted using the analytical formula (\ref{46}). The superposed dots
show the results for the eigenvalues obtained by the numerical
integration of Eq.(\ref{wave}). The agreement is excellent, confirming
the expectation of Appendix A. Table 1 shows, however, that the
agreement with numerical integration becomes poor for the ground state
as its energy approaches zero for large values of $\tilde{a}_2$. This
limitation is understandable from Eq.(\ref{apex2}) derived in Appendix
A and the discussion that follows it.

In summary, we have argued in this paper that it is
legitimate to use the effective range formalism for higher partial
waves in the presence of a coupling to a Feshbach resonance, even though these
parameters may not exist for a long range $1/r^6$ interatomic
potential. Next, we have derived an equation (\ref{46}),
valid {\it under certain restrictions} for any partial wave, relating
the eigenenergies to the effective range parameters.   
This is known to be be valid for $l=0$ and $l=1$ as long as the
range of the potential is much shorter than the oscillator length
We have shown in this paper that it is also applicable for $l=2$, 
so long as the ground state energy is not too close to zero.
Our Eq.(\ref{apex2}), derived for the matching distance for $l=2$
shows the limitation of the shape-independent parameters. At
resonance, since both the terms on the RHS vanish, the analytical 
result (\ref{46}) becomes of limited validity. For $l>2$, the
restrictions are more severe.   


We are indebted to Dr.Takahiko Miyakawa for many useful discussions
at Tokyo University of Science (TUS), where much of the work was
done. We would also like to thank the referee for the incisive
comments that resulted in our adding Appendix A, and a deeper
understanding of the limitations of universality for $l\geq 2$.  
This work was supported by NSERC (Canada), and a grant from
(TUS).
\section{Appendix }
\subsection{Appendix A}
The energy spectrum Eq.(\ref{46}) resulted when the ratio
$c'_2/c'_1$ from (\ref{35}) with the
oscillator potential present was equated to the corresponding ratio from the
scattering solution (\ref{32}) without the confining potential. In
doing so, we assumed that this matching was done at a small enough
distance (still larger than the range $b$ of $V_s(r)$) that the
oscillator potential could be neglected. In this appendix, we justify
this assumption by taking the specific example of $l=2$.
To do this, we need to write Eq.(\ref{harmo1}) in more detail, and
examine the neglected terms in the expansion of Eq.(\ref{35}). For
$l=2$, for small $x$, this is given by
\begin{widetext}
\beq
u_2(x)=c'_1\left[x^3-\frac{\eta_2}{4} x^5+....+\frac{c'_2}{c'_1}
\left(x^{-2}+\frac{\eta_2}{6}+\frac{1}{24}(\eta_2^2-6)
  x^2+...\right)\right]~.
\label{a1}
\eeq
\end{widetext}
It is in the coefficient $\frac{(\eta_2^2-6)}{24}$ of the $x^2$ term that the
oscillator presence is felt; without the oscillator, this coefficient
is $\frac{\eta_2^2}{24}$. In Eq.(\ref{harmo1}), we neglect this term
of order $x^2$, and yet retain the leading term $x^3$ of the regular
solution, even though $x$ is small. This can only be justified if the
matching distance $x$ satisfies the inequality
\beq
x^3 \gg|\frac{c'_2}{c'_1}|\frac{x^2}{4}~.
\label{apex1}
\eeq
To check this, we substitute above the expression (\ref{42}) for
$c'_2/c'_1$, which, for $l=2$, gives the condition
\beq
x\gg \frac{45}{4}\frac{|\tan\delta_2|}{(\sqrt{2}kL)^5}~.
\eeq
To estimate the RHS, we use Eq.(\ref{range}), and put
$k^2=\frac{M}{\hbar^2}E_2$. We then get, for the condition
(\ref{apex1})
\beq
x\gg 2|\left(-\frac{1}{\tilde{a_2}}+\frac{1}{2}\tilde{r}_2
\frac{E_2}{\hbar\omega}\right)^{-1}|
\label{apex2}
\eeq
From Figs.3 and 4, we see that $\frac{1}{\tilde{a_2}}$ is of the order
of $10^{4}$ (either sign), and $\tilde{r}_2$ is $-15\times 10^{3}$ or
ten times larger. The energy $E_2/\hbar\omega$ is of order unity.
Unless there is some accidental cancellation , the
RHS of Eq.(\ref{apex2}) is  very small, of the order of $10^{-4}$. On
the LHS of Eq.(\ref{apex2}), $x$ stands for the matching distance (in
units of $L$, which is slightly larger than
$\tilde{b}=1/15$. Therefore the inequality condition (\ref{apex2}) is
easily satisfied when the scattering length $\tilde{a}_2$ is small. 
This is not the case at resonance, however, when $\tilde{a}_2$ tends
to infinity, and $E_2$ for the ground state approaches zero. We then
see that the RHS of Eq.(\ref{apex2}) tends to infinity, and the
inequality condition cannot be satisfied. Our formula (\ref{46}) is no 
longer accurate for the ground state in this situation. This is 
confirmed by our numerical calculation as shown in Table 1.

\subsection{Appendix  B}
Consider an attractive square-well potential of depth $V_0$ and range
$b$. Define the strength parameter
\beq
s=(\sqrt{MV_0/\hbar^2}~)~b \;.
\label{A1}
\eeq
In a given
partial wave $l$, the expressions for the scattering length $a_l$ and
the effective range $r_l$ are given by
\beq
a_l=-\frac{b^{2l+1}}{(2l-1)!!(2l+1)!!}\frac{j_{l+1}(s)}{j_{l-1}(s)} \;,
\label{A2}
\end{equation}
\begin{widetext}
\begin{equation}
r_l=\frac{(2l-1)!!(2l+1)!!}{b^{2l-1}}\biggl[-\frac{1}{2l-1}
+\frac{2l+1}{s^2}\frac{j_{l-1}(s)}{j_{l+1}(s)}-
\frac{1}{2l+3}\left(\frac{j_{l_1}(s)}{j_{l+1}(s)}
\right)^2\biggr] \;.
\label{A3}
\end{equation}
\end{widetext}


\newpage

\end{document}